\documentclass[aps,apl,twocolumn,reprint,groupedaddress]{revtex4-1} 
\usepackage{verbatim} 
\usepackage{float} 
\usepackage{amsbsy} 
\usepackage{graphicx}
\usepackage{amsmath}

\begin{document}

\title{3D-printed phase waveplates for THz beam shaping}

\author{J. Gospodaric\textsuperscript{1}, A. Kuzmenko\textsuperscript{1}, Anna Pimenov\textsuperscript{1}, C.
Huber\textsuperscript{3}, D. Suess\textsuperscript{3}, S. Rotter\textsuperscript{2}, A. Pimenov\textsuperscript{1} }

\affiliation{\textsuperscript{1}Institute of Solid State Physics, Vienna University of Technology (TU Wien), 1040 Vienna, Austria}

\affiliation{\textsuperscript{2}Institute for Theoretical Physics, Vienna University of Technology (TU Wien), 1040 Vienna,
Austria}

\affiliation{\textsuperscript{3}Christian Doppler Laboratory for Advanced Magnetic Sensing and Materials,
Faculty of Physics, University of Vienna, 1090 Vienna, Austria}

\date{\today} 
\begin{abstract} 
	The advancement of 3D-printing opens up a new way of constructing affordable custom terahertz (THz)
	components due to suitable printing resolution and THz transparency of polymer materials. We present a way of calculating,
	designing and fabricating a THz waveplate that phase-modulates an incident THz beam ($\lambda_{0}=2.14$ mm) in order to create a
	predefined intensity profile of the optical wavefront on a distant image plane. Our calculations were performed for two distinct
	target intensities with the use of a modified Gerchberg-Saxton algorithm. The resulting phase-modulating profiles were used to model the
	polyactide elements, which were printed out with a commercially available 3D-printer. The results were tested in an THz
	experimental setup equipped with a scanning option and they showed good agreement with theoretical predictions.
\end{abstract}
\maketitle

Terahertz spectroscopy is a powerful tool for imaging and testing applications~\cite{Tonouchi2007, Jansen2010}. In analogy to the
situation in optics, elements to manipulate the intensity and shape of a beam are of crucial importance. An efficient way to
design such elements is to combine the holographic principles with computer-supported calculations of the beam
propagation~\cite{Xing_book, Leseberg1992, Matsushima2005}. Such an approach allows not only to manipulate a beam, but moreover to
generate pictures of real and imaginary objects on short time scales. Recently, optical meta-surfaces~\cite{Walther2012, Sun2013,
Yu2014, Genevet2015} added several interesting new ideas to the field, especially on the way to miniaturize the optical units.

Alternatively, optical elements for picture generation may be produced based on phase control of light only. Such elements are
often called phase holograms and they manipulate the incident wave by coordinate-dependent phase changes~\cite{Upatnieks1970,
Gale1993, Lin2014, Zeng2017, Rotter2017}. Well-known examples of phase holograms are dielectric Fresnel lenses or dielectric
diffraction gratings. All such elements can be designed for frequencies as low as terahertz because the underlying Maxwell
equations of light do not contain the wavelength as a parameter.

The goal of this article is to demonstrate that a very practical way to produce such phase holograms is to work with a
commercially available 3D-printer. Additive manufacturing with 3D-printers has gained significant attention in the recent years
due to its versatility, accessibility and generally because it presents a quick, easy-to-use and cost-efficient technique to
produce complex and high-precision structures. The layer height resolution of 3D-printers ($\sim 0.1$~mm) and a high transparency
of the polymers used in 3D-printers \cite{Busch2014} are ideally suited for printing devices that manipulate electromagnetic
radiation in the THz region. Recent examples of 3D printed THz elements are diffraction grating and lenses~\cite{Squires2015}.
Here we make a significant further step in the field of THz imaging by designing and 3D-printing dielectric plates which are able
to reproduce the picture of an arbitrary object.

In our proof-of-principle demonstration we modulate of the phase profile of an incident asymmetric Gaussian beam
($\lambda_{0}=2.14$ mm) at $z_{0}=0$ mm with initially uniform zero phase profile (plane wave) in order to produce two distinct
target intensity profiles (Fig.1(a,b)) at the image plane at $z_{i}=50$ mm, where the $z$-axis represents the optical axis. These
two target profiles in the shape of a cross and of our university logo were represented on a cartesian grid in a circular shape
with a diameter of $53$ mm. Equally spaced grid points were used with a nearest-neighbor distance of $0.53$ mm.
The incident beam was measured by scanning its intensity in the empty channel and was found to have a slightly asymmetric Gaussian profile with
the widths of $13.4$ and $16.6$~mm along the $x$ and $y$-axis of the optical system, respectively. These parameters were then used for
calculating the profile of the phase plates explained below.

\begin{figure*}[tbp]
	\centering{}\includegraphics[width=17cm]{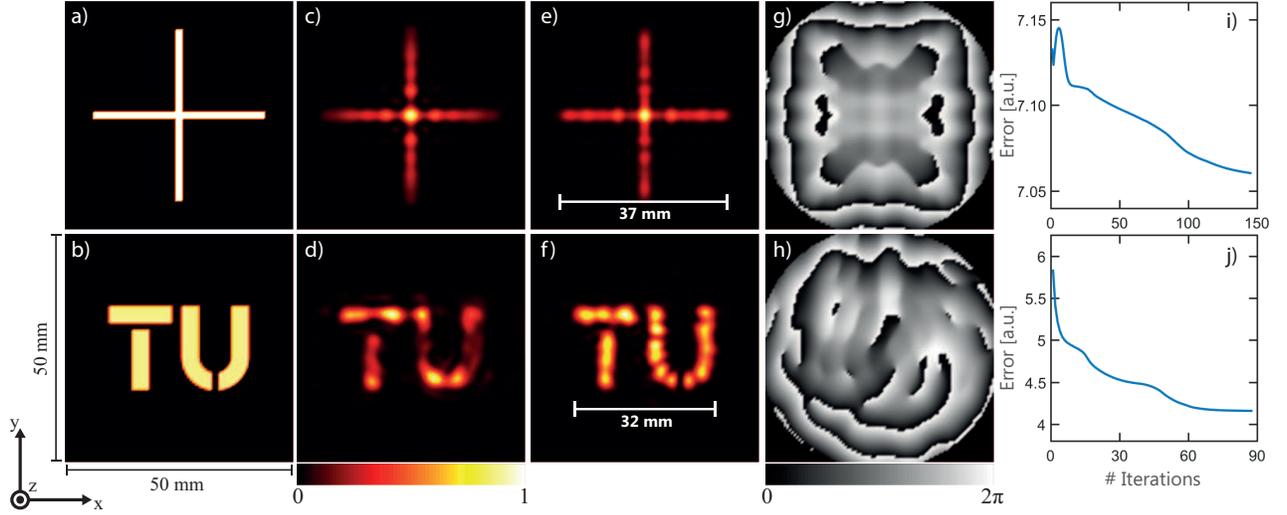}\caption{(a,b) The target intensity profiles 
	used in present experiments. (c,d) Calculated intensity profiles at $z_{i}=50$ mm after the first iteration step of the
	GS algorithm as compared to the result of the last (e,f) iteration step, respectively. (g,h) Calculated phase modulation profiles
	at $z_{0}=0$ mm. (i,j) Algorithm error (mean square deviation from the target profile) as function of the iteration number.}
\end{figure*}

The required phase deformation of the incident beam was calculated using the Gerchberg-Saxton (GS) iterative algorithm
\cite{Gerchberg1972}. This algorithm determines the phase of the optical wave function in the imaging system, whose intensity in
the diffraction (source) and the image plane of the system are known. The method typically assumes a Fourier Transform relation 
between the optical fields on both planes. Therefore, in order to move the image plane from far-field to a finite distance from the 
diffraction plane we considered a propagation function that linked the optical fields on both
 planes based on the Huygens principle. In this spirit the wavefront on the source plane,
$E^{s}$, was represented as point emitters of spherical wavelets with respective coordinates $\left(x,y\right)$, amplitude
$A_{\left(x,y\right)}$ and phase angle $\varphi_{\left(x,y\right)}$. The optical field on the distant plane, $E^{d}$, was then
calculated by summing up all contributions from every point emitter on $\left(x,y\right)$ for every point $(x',y')$ of the grid on
the distant plane as:

\begin{align}
\nonumber	E_{(x,y)}^{s}\left(x',y'\right)=\frac{A_{(x,y)}}{r_{(x,y),(x',y')}}e^{-i\mathbf{k}\mathbf{r}_{(x,y),(x',y')}+i\varphi_{(x,y)}},\\
		E_{(x',y')}^{d}=\sum_{(x,y)}E_{(x,y)}^{s}\left(x',y'\right)\left[1+\cos\left(\Omega\right)\right], 
\end{align}

where $\mathbf{r}$ represents the vector between the two points on different planes, $\boldsymbol{k}$ is the wave vector and $\Omega$ is the
angle between the normal of the source plane and the vector $\boldsymbol{r}$. 

Equation~(1) corresponds to the Fresnel-Kirchhoff diffraction formula \cite{Born1999}. Due to the loss of the amplitude information of the optical field
profiles during the GS algorithm, the pre-factors in Eq.~(1) were left out. Note also that the part in square brackets represents
the Fresnel inclination factor. 

As the GS algorithm utilizes a subsequent back and forth propagation of the beam, the roles of the two image and diffraction 
planes at $z_{0}$ and and $z_{i}$ were interchanged in Eq.~(1). We note that the reverse propagation requires the change 
of sign in the exponent in Eq.~(1).

Figure~1(c,d) shows the calculated intensity profiles in the image plane for two targets after the first iteration step of the
algorithm. Compared to the results after the final convergence of the algorithm (Fig.~1(e,f)) it can be seen that already the
first iteration step produces reasonable intensity profiles. The calculated mean square deviations shown in Fig.~1(i,j)
demonstrate that final convergence of the GS algorithm is reached after about $\sim 100$ iterations which takes about 4 hours of
computer time on a commercial laptop. Analyzing the evolution of the errors with increasing iteration steps, we note that the
relative improvement is rather small. This fact is due to the large amount of the zero intensity points in the
target, which masks the advancement of the image.

\begin{figure}[tbp] 
	\centering{}\includegraphics[width=8.5cm]{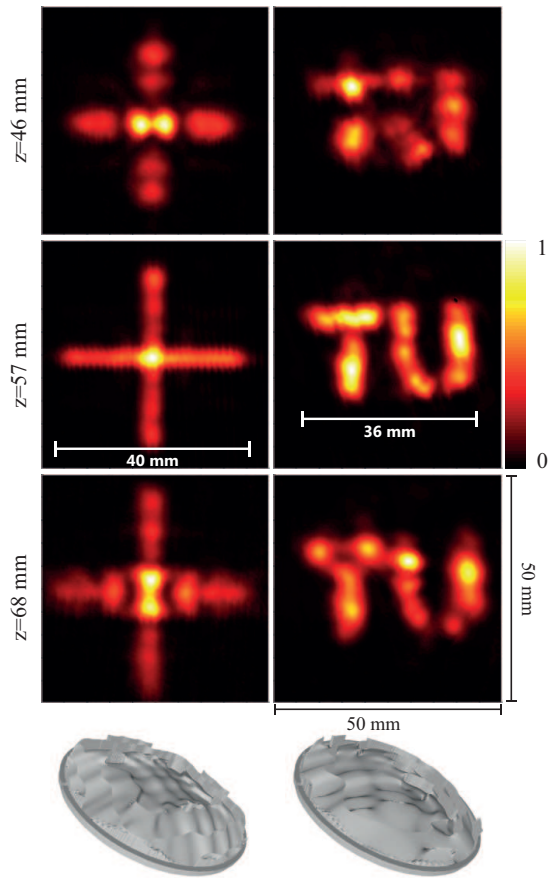}\caption{Experimental intensity profiles of the
	optical field produced by an incident THz beam that was phase modulated by the two printed elements. The images
	appeared the sharpest at $z=57$ mm. Bottom panels show computer models of the phase plates used in the experiment that were 
	modeled according to Fig. 1(g,h) and designed to fit a lens holder.} 
\end{figure}

Based on the calculated phase profiles shown in Fig. 1(g,h) as gray scales, two optical elements, corresponding to the two
targets, were modeled by using the formula $z_{\left(x,y\right)}={\varphi_{\left(x,y\right)}\lambda}/{[(n-1)2\pi]}$ to calculate
the thickness of the optical elements at transverse positions $\left(x,y\right)$ in order to achieve the correct phase-shift
profile. The refractive index $n=1.52\pm 0.03$ of polyactide (PLA) - the plastic lament of the commercial additive 3D-printer -
was measured in a separate experiment at the operating wavelength. This material was used to print the elements with a nozzle
diameter of $0.4$ mm and spatial resolution of $0.1$ mm. The printed elements were positioned into the path of the THz beam at the
$z_{0}$-value of the optical system. The incident beam was produced by an IMPATT diode with 30 mW power. The transverse profile
intensity of the phase-modulated optical field was then measured by using a pyroelectric detector on a translator in the
$xy$-plane, where a sufficient signal-to-noise ratio was achieved by the implementation of a lock-in detection system. A sequence
of intensity profiles was measured at various positions around the set image plane position $z_{i}$. In Fig. 2 we present the
measured intensity profiles at three different positions $z$. We found that the sharpest images, which were the closest to the
simulated profiles in Fig. 1(e,f), were located at $z=57$ mm, slightly deviating from the expected distance of $50$ mm.
Simultaneously, the effective shape size of the images is also increased compared to respective target images shown in Fig.
1(e,f) by approximately $10$ \%.

In order to find the reason for the variation of image size we investigated the influence of the phase distribution of the
incident beam. Indeed, in the real experiment the phase profile of the beam deviates from an ideal plane wave and is better
approximated with a spherical beam with a large but yet unknown curvature $R$. Assuming a finite wavefront radius for the phase
part of the optical field of the Gaussian incident beam causes a displacement of the image plane at which the shape of the
intensity profile appears the sharpest. In addition, the size of the image increases proportionally to the change in the optimal
distance (see Fig. 3(b,c)). 

For each value of $R$ a set of intensity profiles at various positions $z$ with a step size of $\Delta
z=1$ mm was simulated. From it we were able to select the position $z$, at which the shape in the simulated intensity profile
appeared focused, for each value of $R$ (see Fig. 3(b)). Figure 3 also shows that the dependence of image plane position on $R$
coincides with the simple thin lens equation with the focus of $z_{i}$ (depicted in black), in particular in the linear regime of
the equation ($\frac{R}{z_{0}}\gg1$). Effective sizes of the sharpest shapes in respect with $R$, shown in Fig. 3(c), follow the
same trend as their positions. This may be expected as a result of a linear dependence between the image plane position of the
sharpest image and its position on the optical axis. For $R=500$ mm we see in Fig. 3(b) that the sharpest image appears at
$z=56$ mm, relatively close to the position of the sharpest measured image in Fig. 2. In addition, the effective size of the
sharpest image for $R=500$ mm is comparable to the measured one and thus shows that an uncollimated beam with a similar curvature
explains well the mismatch of the theoretical prediction and the experimental results, despite our optimization of the optical
system.

We also investigated the effect of varying the wavelength of the incoming beam keeping the phase plate fixed the one producing the
cross-shaped intensity pattern (see Fig.~1(g)). Simulated intensity profiles at the image plane $z_{0}$ for two wavelengths around
the value $\lambda_{0}$ are shown in Fig. 3(a). After comparing these results to Fig. 1(e) we conclude that a noticeable change of
wavelength still produces intensity profiles that preserve the basic shape of the target yet it also shows that a specific
spectral width of the radiation would influence the sharpness of the image.

\begin{figure}[tbp] 
	\centering{}\includegraphics[width=8.5cm]{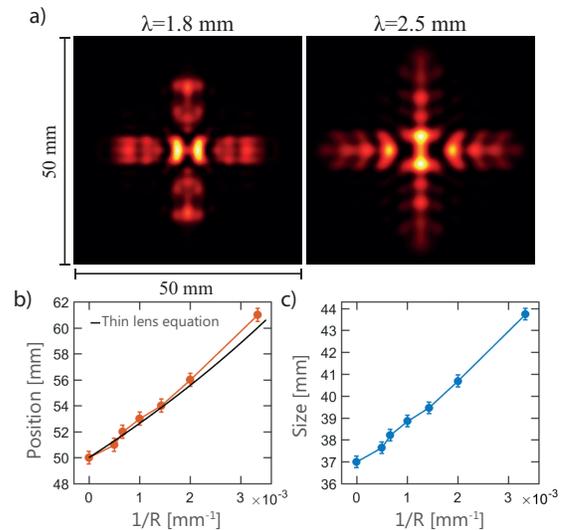}\caption{a) Simulated intensity profiles of the
	optical wave function at the image plane, where the wavelength was set to $1.8$ and $2.5$ mm, respectively, and the incident beam
	was modulated by the waveplate in Fig. 1(g), optimized for $\lambda = 2.14$~mm. (b,c) Position and size of the image as function
	of the wavefront curvature of the incident beam ($R$) including the values from Fig. 1, where $R=\infty$.
	Points - simulations, solid black line represents the thin lens formula
	with a focus set at $z_{i} = 50$~mm.} 
\end{figure}

We also mention parenthetically that we improved the propagation function in Eq. (1) by better taking into account the finite
thickness of the phase plate. Instead of an instant change of the propagation phase (as for a waveplate with zero thickness), a
finite height has been added to each emitter of the source plane with a value corresponding to the actual thickness at a point
(x,y). This improvement slightly corrected the simulated intensity profiles when compared to real results, yet the changes are
barely seen directly and are therefore not shown here.

In conclusion, we demonstrate that versatile wave plates for THz beam shaping can be very easily produced with a commercially
available 3D printer. Specifically, we showed how such dielectric phase plates can modulate an incoming Gaussian beam such that it
produces an arbitrary image on the detector plane. Our results are expected to facilitate THz imaging on all levels where
customized and cost-efficient beam shaping solutions are required.

The financial support by the Austrian Federal Ministry of Science, Research and Economy, by the
National Foundation for Research, Technology and Development 
and by the Austrian Science Funds (W1243) is acknowledged.

\nocite{*}

\end{document}